\documentclass[twocolumn,showpacs,amsmath,amssymb,superscriptaddress]{revtex4-1}
\usepackage{amsmath,amssymb,color,latexsym,natbib,graphicx}
\usepackage{dcolumn}
\usepackage{bm}
\usepackage{color}
\usepackage{ulem}
\usepackage{soul}

\def\be{\begin{equation}}
\def\ee{\end{equation}}
\def\ba{\begin{eqnarray}}
\def\ea{\end{eqnarray}}

\begin{document}
\preprint{1}

\title{Nonlinear Diffusion Models for Gravitational Wave Turbulence}
\author{S\'ebastien Galtier}
\affiliation{Laboratoire de Physique des Plasmas, Univ. Paris-Sud, Universit\'e Paris-Saclay, \'Ecole polytechnique F-91128 Palaiseau Cedex, France}
\affiliation{Institut universitaire de France}
\email{sebastien.galtier@u-psud.fr}
\author{Sergey V. Nazarenko}
\affiliation{INPHYNI, CNRS, Universit\'e C\^ote d'Azur, France}
\email{sergey.nazarenko@inphyni.cnrs.fr}
\author{\' Eric Buchlin}
\affiliation{Institut d'Astrophysique Spatiale, b\^at.\,121, CNRS, Univ. Paris-Sud, Universit\'e Paris-Saclay, F-91405 Orsay, France}
\email{eric.buchlin@ias.u-psud.fr}
\author{Simon Thalabard}
\affiliation{Instituto Nacional de Matem\'atica Pura e Aplicada, IMPA, 22460-320 Rio de Janeiro, Brazil}
\email{simon.thalabard@impa.br}
\date{\today}

\begin{abstract}
A fourth-order and a second-order nonlinear diffusion models in spectral space are proposed to describe gravitational wave turbulence in the approximation of 
strongly local interactions. We show analytically that the model equations satisfy the conservation of energy and wave action, and reproduce the power law 
solutions previously derived from the kinetic equations with a direct cascade of energy and an explosive inverse cascade of wave action. In the latter case, we show 
numerically by computing the second-order diffusion model that the non-stationary regime exhibits an anomalous scaling which is understood as a self-similar 
solution of the second kind with a front propagation following the law $k_f \sim (t_*-t)^{3.296}$, with $t<t_*$. These results are relevant to better understand the 
dynamics of the primordial universe where potent sources of gravitational waves may produce space-time turbulence. 
\end{abstract}

\maketitle

%Keywords: Wave turbulence -- Diffusion model -- Inverse cascade -- Cosmology

%%%%%%%%%%%%%%%%%%%%%%%%
\section{Introduction}

The nonlinear nature of the Einstein equations of general relativity suggests that space-time can be turbulent. Such a turbulence has been studied 
in the context of spinning black holes \cite{adams14,green14,yang15b,yang15} by using the gravity-fluid correspondence. It is shown numerically that 
such a system can display a nonlinear parametric instability with transfers reminiscent of an inverse cascade; the precise mechanism is, however, not 
totally understood. The possibility of having a turbulent cascade within the metric perturbations was already discussed in the past \cite{Efroimsky94} but 
it is only recently that a rigorous theory has been proposed for the regime of gravitational wave (GW) turbulence \cite{GN17}. The presence of small 
nonlinearities has been exploited to derive a four-wave kinetic equation which describes the spectral transfers of energy and wave action. In other words, 
the theory explains the nonlinear evolution of weak ripples on the Poincar\'e-Minkowski flat space-time metric. The theory is limited to a $2.5+1$ 
diagonal metric tensor \cite{Hadad2014} which includes only one type ($+$) of GW (the $\times$ GW is excluded). Besides the kinetic equations, the 
main results obtained are the derivation of its power-law (constant flux) solutions and the demonstration that we have a direct energy cascade and an 
explosive inverse cascade of wave action (a property of finite-capacity turbulence systems) with a priori the possibility to excite Fourier modes from the 
injection wavenumber 
$k_I$ to $k=0$ in a finite time. However, and as discussed in \cite{GN17}, such a transfer driven by GW turbulence stops at an intermediate scale 
where the turbulence regime becomes strong. Note that the change of regime does not preclude the possibility to extend such an inverse cascade to $k=0$ 
in a finite time \cite{GNL18}. Generally speaking, we may say that space-time turbulence is likely to be a relevant regime for describing the very early 
universe soon after the Planck's time when the universe emerges from a quantum foam \cite{Wheeler1955}. 

The main goal of the present paper is to  study the properties of GW turbulence further and to compare it with other turbulent systems. The kinetic equation
derived in \cite{GN17} is, however, too complicated for detailed analytical or numerical studies. Therefore, our strategy is to derive nonlinear diffusion 
models (also called differential models) which correspond to strongly local interaction approximations of the kinetic equation. This type of reduction, first introduced 
by Leith in 1967 \cite{Leith67} to study three-dimensional Navier-Stokes turbulence, is quite common and plays an important role for achieving a qualitative 
and even quantitative understanding of turbulence in various physical situations, for both weak and strong turbulence \cite{Dyachenko,Zakharov99,GB10}. 
For example, the anomalous scaling of the non-stationary spectrum, a property of finite-capacity turbulence systems first detected in the numerical simulations 
of some kinetic equations \cite{galtier00,lacaze01}, has been easily studied in other systems using differential models 
\cite{CN04,N06,LN06,Boffetta09,Proment12,Thalabard15}. In particular, it was established that the anomalous exponent is independent of the initial conditions 
but varies for different models of the same physical system. 

In the present paper, we propose two diffusion models for GW turbulence: a fourth-order and a second-order model which are introduced in sections 
\ref{part2} and \ref{part3}, respectively. Their derivation is based, in particular, on the phenomenology of wave turbulence that was introduced in \cite{GN17}. 
Numerical simulations of the second-order diffusion model are then performed to study the form of the front propagation during the inverse cascade 
of wave action. The results are presented in section \ref{part4}. Conclusions are discussed in section \ref{part5}.

%%%%%%%%%%%%%%%%%%%%%%%%
\section{Fourth-order diffusion model} \label{part2}

Nonlinear diffusion models have proved to be a very useful tool in the analysis of both strong \cite{Leith67,CN04} and wave turbulence 
\cite{hasselmann,Zakharov99,Boffetta09,GB10}. Here, we shall derive such a model for GW turbulence. Since the leading nonlinear interaction of GW is 
four-wave interaction \cite{GN17}, the model is a fourth-order diffusion equation of the type
\be
\frac{\partial N(k)}{\partial t} = \frac{\partial^2}{\partial k^2} \left[ D N^4(k) \frac{\partial^2 (k^2/N(k))}{\partial k^2} \right] \, ,
\label{E1}
\ee
where $N(k)$ is the one-dimensional wave action spectrum, $k$ the wavenumber and $D$ a diffusion coefficient. This equation is constructed 
in such a way that it preserves the nonlinearity degree with respect to the spectrum (cubic in our case) and, its cascade and thermodynamic solutions. 
It is only in rare situations that one can derive a diffusion equation directly from the kinetic equation of wave turbulence (by taking the strongly local 
interactions limit). These exceptions concern nonlinear optics \cite{Dyachenko} and magnetohydrodynamics \cite{GB10}. In our case, we will also 
use the phenomenology of wave turbulence \cite{GN17}. A dimensional analysis of expression (\ref{E1}) gives the following information about $D$
\be
\frac{N(k)}{\tau} \sim \frac{D N^3(k)}{k^2} \, , 
\ee
and thus
\be
D \sim \frac{k^2}{\tau N^2(k)} \, .
\ee
We also have the scaling relation \cite{Maggiore08,GN17}
\be
E(k) \sim h^2 k \sim \omega N(k) \sim k N(k) \, , 
\ee
where $E(k)$ is the one-dimensional energy spectrum, $h$ the amplitude of the metric perturbation (ie. $g_{\mu\nu}=\eta_{\mu\nu} + h_{\mu\nu}$, 
with $\eta_{\mu\nu}$ being the Minkowski metric and $h \sim h_{\mu\nu}$) and $\omega=kc$ with $c$ being the speed of light; thus $N(k) \sim h^2$. This gives
\be
D \sim \frac{k^2}{\tau h^4} \sim \frac{k^2}{(\tau_{GW} / \epsilon^4) h^4} \sim \frac{k^2}{\tau_{GW}} \sim k^3 \, ,
\label{diff1}
\ee
where the GW time is given by the relation $\tau_{GW} \sim 1/ \omega$ and where 
$\epsilon \sim h_{\mu\nu}/\eta_{\mu\nu} \sim h \ll 1$ is a small parameter. 
The introduction of expression (\ref{diff1}) into (\ref{E1}) leads to the 
following fourth-order diffusion equation for isotropic three-dimensional GW turbulence
\be
\frac{\partial N(k)}{\partial t} = A \frac{\partial^2}{\partial k^2} \left[ k^3 N^4(k) \frac{\partial^2 (k^2/N(k))}{\partial k^2} \right] \, ,
\label{equadif}
\ee
where $A$ is a positive constant (presumably of order one). 

Equation (\ref{equadif}) conserves the wave action $\int N(k) dk$ and energy $\int \omega N(k) dk$. Indeed, if we define 
\be
\frac{\partial N(k)}{\partial t} = \frac{\partial^2 K(k)}{\partial k^2}  = - \frac{\partial Q(k)}{\partial k} \, ,
\ee
with $Q(k)$ being the wave action flux, then (for simplicity $c=1$)
\ba
\frac{\partial E(k)}{\partial t} &=& k \frac{\partial N(k)}{\partial t} = - \frac{\partial}{\partial k} \left[ K(k) - k \frac{\partial K(k)}{\partial k} \right] \nonumber \\
&=& - \frac{\partial P(k)}{\partial k} \, , 
\label{equE}
\ea
with $P$ being the energy flux. 
We can check that the thermodynamic (zero flux) solution $N(k) \sim k^2T/(k+\mu)$ \cite{GN17} is satisfied by equation (\ref{equadif}). 

We can also find the constant (non-zero) flux solutions and find  the cascade directions. Let us introduce $N(k)=C_N k^{\alpha}$ into 
equation (\ref{equadif}); after simple calculations we obtain 
\be 
Q(\alpha) = - (1-\alpha)(2-\alpha)(3+3\alpha) k^{2+3\alpha} A C_N^3 \, .
\ee
Therefore, a constant wave action flux solution corresponds to $\alpha=-2/3$. For this value we find $Q_0 \equiv Q(-2/3) = -(40/9) A C_N^3<0$, 
which means that this solution corresponds to an inverse cascade (the wave action spectrum is positive definite thus $C_N >0$). 
Let us now substitute $N(k)=C_E k^{\beta}$ into equation (\ref{equE}); after some calculations we obtain 
\be
P(\beta) = (1-\beta)(2-\beta)(-2-3\beta) k^{3+3\beta} A C_E^3\, .
\ee
The constant energy flux solution corresponds to $\beta=-1$; in this case we have $P_0 \equiv P(-1)=6AC_E^3>0$, which means that we 
have a direct cascade of energy (because $C_E>0$).
This analysis gives a prediction for the Kolmogorov constants $C_Q$ and $C_P$, which depend, however, on $A$. We find
\be
N(k) = (- Q_0)^{1/3} C_Q k^{-2/3} \, , 
\ee
with the Kolmogorov constant $C_Q=(9/(40A))^{1/3}$ and 
\be
E(k) = P_0^{1/3} C_P k^{0} \, ,
\ee
with the Kolmogorov constant $C_P=(1/(6A))^{1/3}$. 
Interestingly, the ratio of the Kolmogorov constants becomes independent of $A$:
\be
\frac{C_Q}{C_P} = \left( \frac{27}{20} \right)^{1/3} \simeq 1.105 \, . 
\ee
Note that a similar situation was also found with a differential model for two-dimensional hydrodynamic turbulence \cite{LN06}.

%%%%%%%%%%%%%%%%%%%%%%%%
\section{Second-order diffusion model}\label{part3}

The nonlinear diffusion model of GW turbulence gets simplified if we do not include the thermodynamic solutions: it becomes a second-order diffusion equation 
which is easier to simulate numerically. In our case, we have
\be
\frac{\partial N(k)}{\partial t} = B \frac{\partial}{\partial k} \left[ k^2 N^2(k) \frac{\partial (k N(k))}{\partial k} \right] \, ,
\label{equadif2}
\ee
where $B$ is a positive constant (presumably of order one). We can check that the constant flux solutions discussed above are recovered by this equation, 
and that the wave action and the energy are conserved (the relation $dP=kdQ$ can be used). By using similar notation as above, we can demonstrate 
that $Q_0 \equiv Q(-2/3)=-(1/3)B C_N^3$ and $P_0 \equiv P(-1)=(1/3)B C_E^3$, which means that the directions of the cascades are recovered. 
For this model we obtain
\be
N(k) = (- Q_0)^{1/3} C_Q k^{-2/3} \, , 
\label{sol1}
\ee
and 
\be
E(k) = P_0^{1/3} C_P k^{0} \, ,
\ee
with equal Kolmogorov constants, $C_Q=C_P=(3/B)^{1/3}$. 

Time-dependent solutions of equation (\ref{equadif2}) will be studied in section \ref{part4}. In particular we shall find a non-stationary solution with the 
wave action spectrum propagating towards small wavenumbers. This system is of finite capacity, ie. the integral 
\be
\int_0^{k_I} N(k) dk \, , 
\ee
is finite when the solution (\ref{sol1}) is used. This leads to an anomalous scaling with a non-trivial power-law. This non-stationary spectrum may 
be modelled as a self-similar solution of the second kind (see eg. \cite{Falkovich91,Thalabard15}) taking the form
\be
N(k) = \frac{1}{\tau^{a}} N_0 \left(\frac{k}{\tau^b} \right) \, , 
\ee
where $\tau = t_*-t$, and $t_*$ is a finite time at which the wave action spectrum reached the smallest available wavenumber. 
By introducing the above expression into (\ref{equadif2}) we find  the condition
\be
2 a-b=1 \, . 
\ee
A second condition can be found by assuming that $N_0(\xi) \sim \xi^m$ far behind the front. Then, the stationarity condition gives the following relation
\be
a + m b = 0 \, . 
\ee
Finally, the combination of both relations gives 
\be
m=-\frac{a}{b} = -\frac{1}{2}-\frac{1}{2b} \, . 
\ee
The latter expression means that we have a direct relation between the power law index $m$ of the spectrum and the law of the front propagation 
which follows $k_f \sim \tau^b$. For example, if we assume that the stationary solution -- the Kolmogorov-Zakharov (KZ) spectrum -- is established 
immediately during the front propagation \cite{Falkovich91}, then $m=-2/3$ and $b = 3$ (and $a=2$). In this case, the prediction for the front 
propagation is 
\be
k_f \sim (t_*-t)^3 \, . 
\ee
Any deviation from this prediction has to be considered as an anomalous scaling which is sometimes difficult to observe numerically because the
deviation is often tiny \cite{Bell2017,Grebenev2017}.

%%%%%%%%%%%%%%%%%%%%%%%%
\section{Numerical simulation}\label{part4}

\begin{figure}
\includegraphics[width=1.0\linewidth]{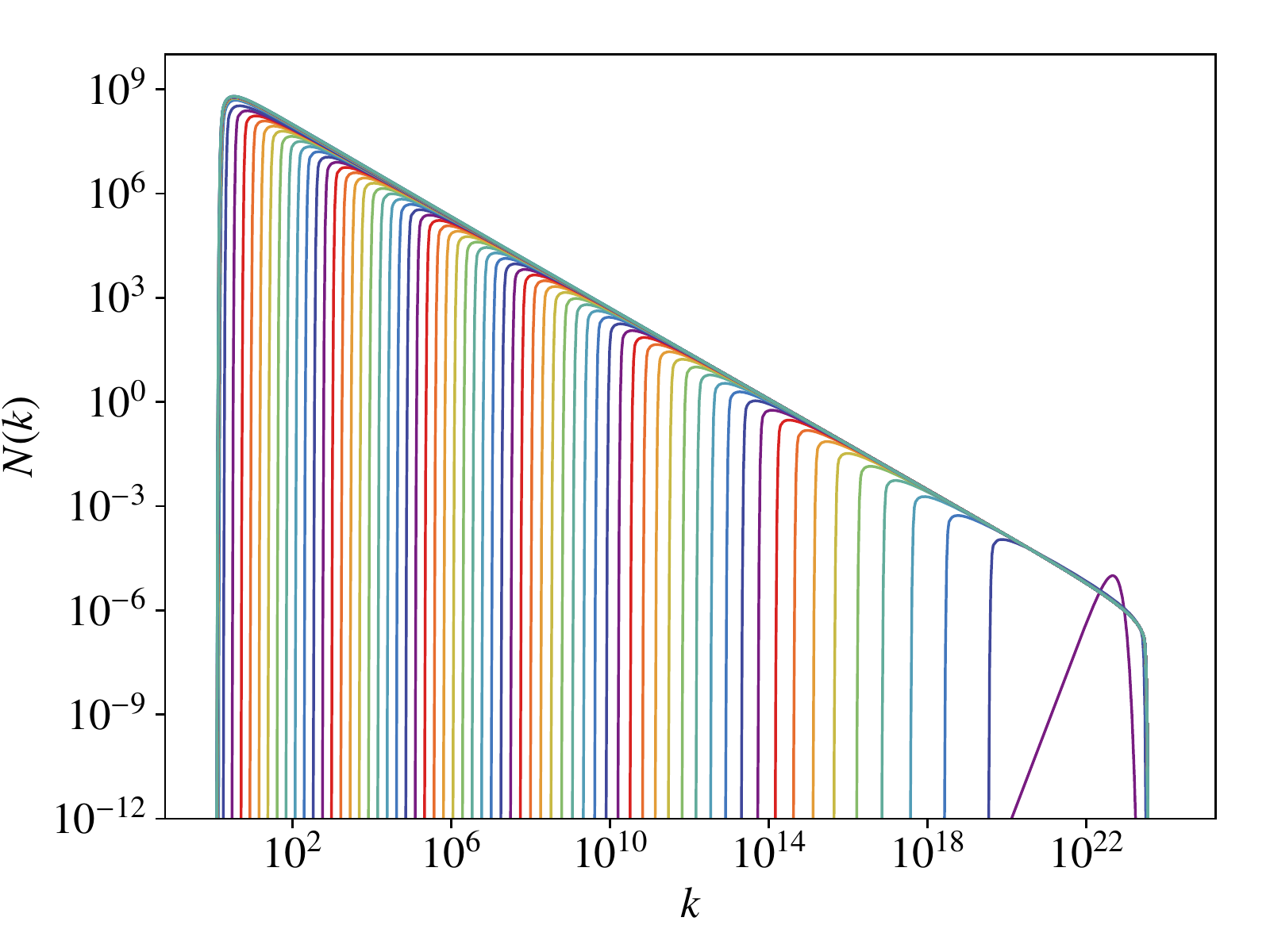}
\includegraphics[width=1.0\linewidth]{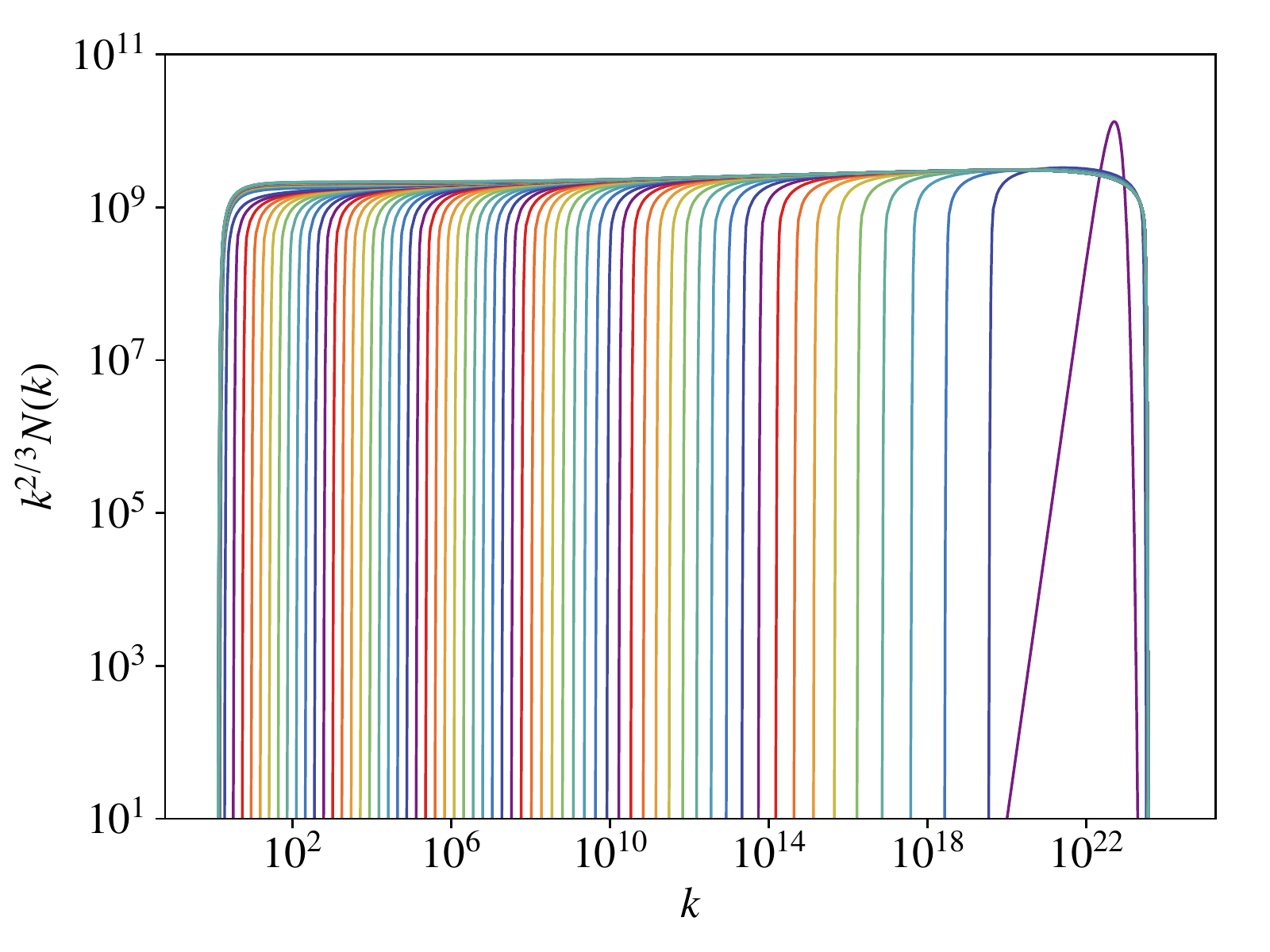}
\caption{Time evolution of the wave action spectrum with (bottom) and without (top) a compensation by $k^{2/3}$. 
The initial spectrum is localized at small scales ($k \sim10^{22}$).}
\label{Fig1}
\end{figure}

In this section we shall investigate numerically the time evolution of the wave action spectrum described by the second-order diffusion equation 
(\ref{equadif2}) with $B=1$. Linear hyper-viscous and hyper-hypoviscous terms are added in order to introduce sinks at small scale and large scale, respectively, 
to avoid the development of numerical instabilities. Then, the following equation is simulated
\ba
\frac{\partial N(k)}{\partial t} =&& \frac{\partial}{\partial k} \left[ k^2 N^2(k) \frac{\partial (k N(k))}{\partial k} \right] \nonumber \\
&-& \nu k^4 N(k) - \eta \frac{N(k)}{k^4} \, ,
\label{equadif3}
\ea
with $\nu = 10^{-85}$ and $\eta = 10^{20}$. A logarithmic subdivision of the $k$-axis is used with $k_i= 2^{i/10}$ and $i$ an integer varying between
$0$ and $799$. A Crank-Nicholson numerical scheme is implemented for the nonlinear term and an adaptive time-step is used. No forcing term is added. 
The code is publicly available from \url{https://git.ias.u-psud.fr/ebuchlin/nldiffus-gw}. 

\begin{figure}
\includegraphics[width=1.0\linewidth]{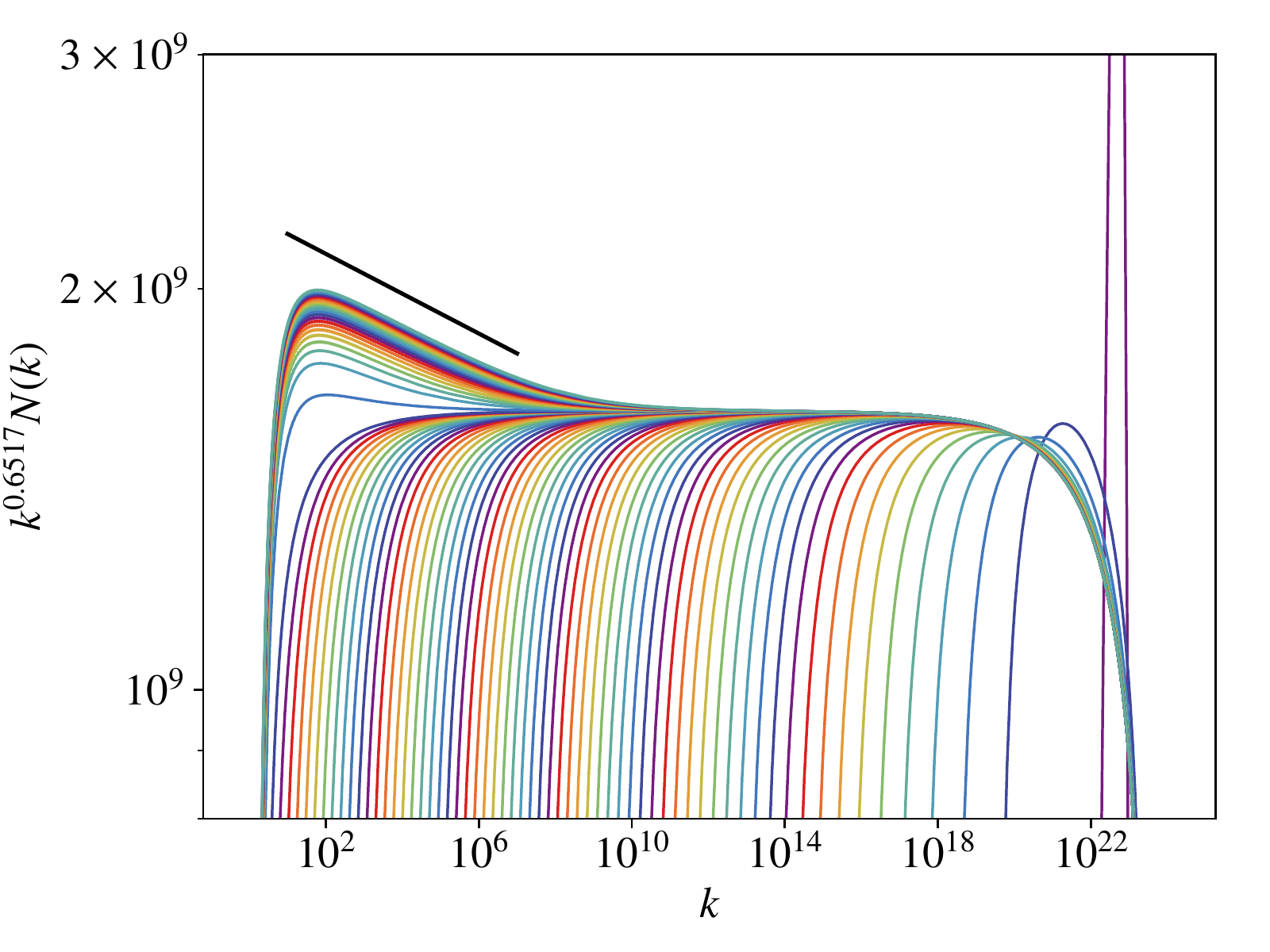}
\includegraphics[width=1.0\linewidth]{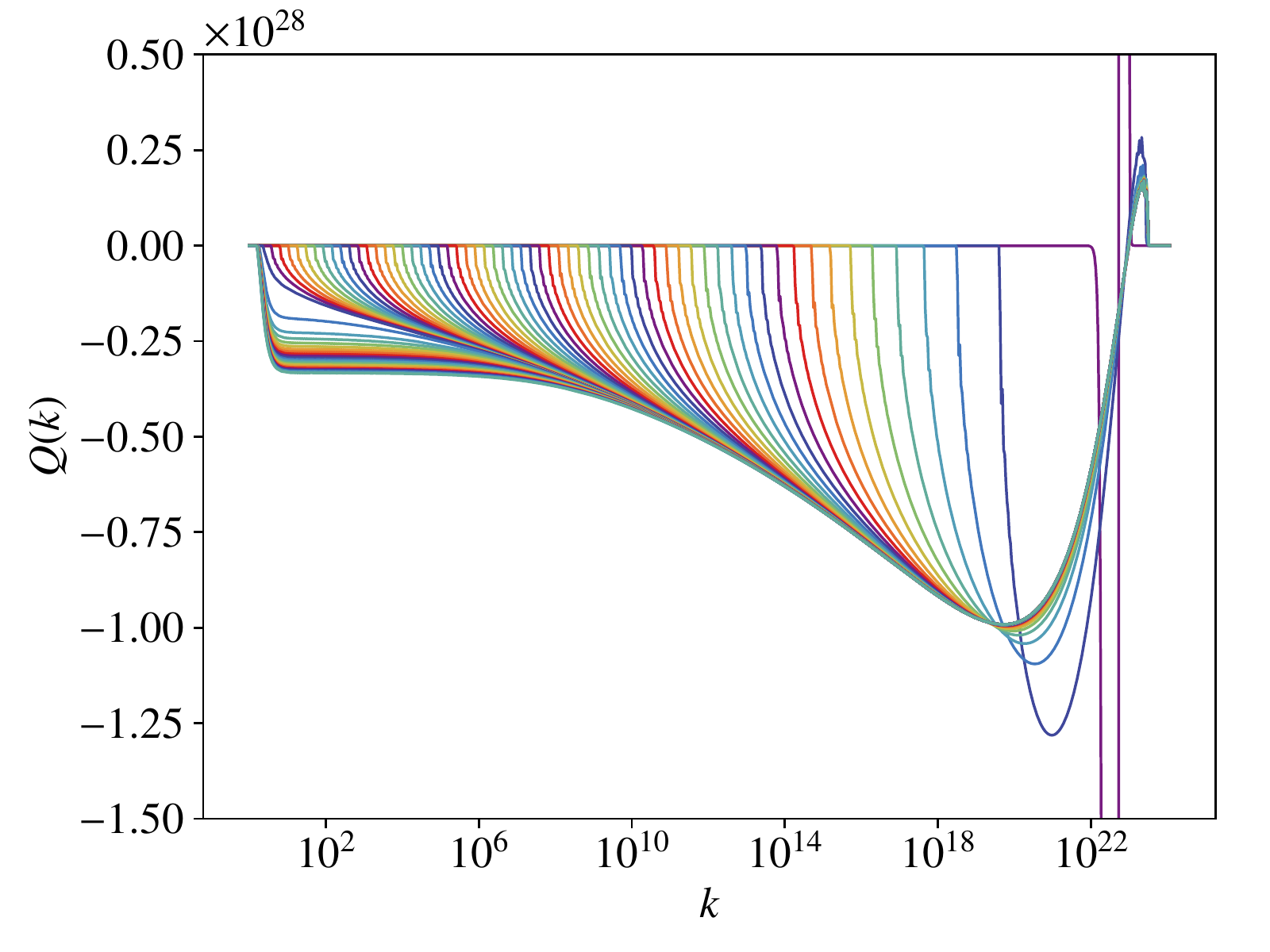}
\caption{Top: time evolution of the compensated (by $k^{0.6517}$) wave action spectrum; the Kolmogorov-Zakharov solution is established only after 
the hyper-hypoviscous scale is reached (the black segment corresponds to a spectrum in $k^{-2/3}$). Bottom: time evolution of the wave action flux $Q(k)$.}
\label{Fig2}
\end{figure}

Figure \ref{Fig1} shows the time evolution (every $6000$ time-steps) of the wave action spectrum with an initial injection at $k_I \sim 10^{22}$. As expected, 
an inverse cascade appears with a spectrum reaching the smallest scale available ($k \sim 1$). At first glance, the KZ solution in $k^{-2/3}$ (over more than 
20 decades!) seems to be formed as we can see at the bottom of Fig.\,\ref{Fig1} where compensated spectra are displayed. 

\begin{figure}
\includegraphics[width=1.0\linewidth]{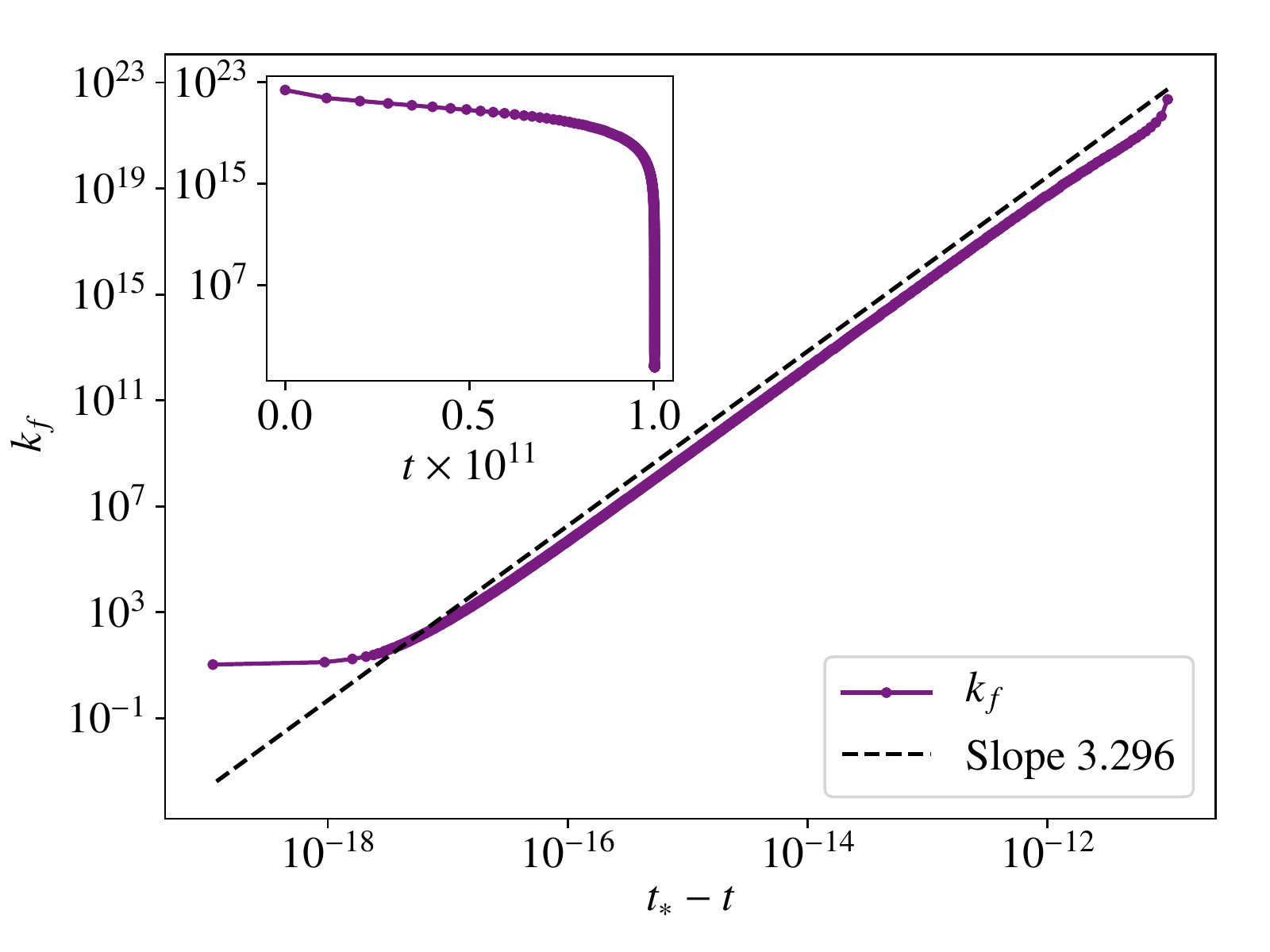}
\caption{Temporal evolution of the spectral front $k_f$ for $t \le t_*$ in linear-logarithmic coordinates (inset) and as a function of $t_*-t$ in double 
logarithmic coordinates. The black dashed line corresponds to $(t_*-t)^{3.296}$.}
\label{Fig3}
\end{figure}
The anomalous exponent can be determined very precisely  by solving numerically  an eigenvalue problem characterizing the existence of self-similar solutions 
of the second kind  to the evolution equation. Such solutions  were described in \cite{Thalabard15}  for a broad class of second-order diffusion equations, exhibiting 
(forward) flux solutions with finite ultra-violet capacity. In the present case, a similar analysis carries through provided one maps  the inverse cascade behavior of 
the evolution equation  (\ref{equadif2})  into a direct cascade in physical space. Explicitly, with the change of variables  $k \to 1/\ell $, $N \to \tilde N = N/\ell ^2$, the 
evolution equation (\ref{equadif2}) becomes 
\be
\frac{\partial \tilde N(\ell)}{\partial t} = \frac{\partial}{\partial \ell} \left[ \ell^4 \tilde N^2(\ell) \frac{\partial (\ell \tilde N(\ell))}{\partial \ell} \right] 
\label{equadif_tilde}.
\ee
This renormalized system has the equilibrium solution $\tilde N \sim \ell ^{-1}$ and the  flux solution $\tilde N \sim \ell ^{-4/3}$, which has finite ultra-violet capacity~: the 
general framework of \cite{Thalabard15} then applies. Using the numerical dichotomic procedure herein described,  we determine the anomalous exponent for the reduced 
system as $m-2 \simeq -1.3483$, that is $m \simeq -0.6517$. This anomalous behaviour is well observed in Fig. \ref{Fig2} (top) which shows the compensated spectra. 
Clearly the KZ solution is established only at $t>t_*$, ie. after the largest (hyper-hypoviscous) scales are reached. This agrees with the time evolution of the wave action flux 
(bottom): a plateau corresponding to the KZ solution appears  at the latest times only. Note that the simulation is stopped before the formation of the KZ spectrum over the 
entire range of scales. 

We may also investigate the front propagation towards small wavenumbers and check if the power-law 
\be
k_f \sim (t_*-t)^{3.296} \,, 
\label{as}
\ee
corresponding to the anomalous scaling $m \simeq -0.6517$ (and also $a \simeq -2.148$ and $b \simeq 3.296$) is observed. For that, one needs to 
follow the front propagation $k_f(t)$ which will be defined by using the compensated spectra in Fig. \ref{Fig1} (bottom): the value $1.4\times10^9$ for 
the compensated spectra is chosen to define the front $k_f(t)$. 
The result (Fig.~\ref{Fig3}; inset) displays a sharp decrease of the wavenumber of the front at a time close to $10^{-11}$, which will be used to define $t_*$. 
We see (Fig.~\ref{Fig3}) that the expected power-law \eqref{as} is well observed over six decades. 
This result illustrates the explosive character of the inverse cascade of wave action in GW turbulence.
Note that the 24 decades in wavenumbers used for the simulation are necessary to detect without ambiguity the anomalous scaling. However, as explained above,
it is expected that GW turbulence becomes strong at large scale, which prevents the formation of such an extended power-law range.

%%%%%%%%%%%%%%%%%%%%%%%%
\section{Conclusion}\label{part5}
In this paper we have proposed two nonlinear diffusion models for GW turbulence which reproduce the properties previously derived from the kinetic equation 
\cite{GN17} (power-law solutions, cascade directions). We have also derived some specific properties like the Kolmogorov constants. Additionally, we have 
performed a numerical simulation of the second-order diffusion model to illustrate the existence of an explosive inverse cascade of wave action, and have 
demonstrated the existence of an anomalous scaling which is typical to finite-capacity turbulence systems. This finding leads to a non-trivial power-law behind 
the propagating front in the inverse cascade spectrum. 
This analysis of the anomalous spectrum in the case of an inverse cascade is the first made with a diffusion model. Note, however, that a similar analysis based 
on kinetic equations has already been done by \cite{Semikoz95,Semikoz97,lacaze01} to study the formation of a Bose-Einstein condensate.

%%%%%%%%%%%%%%%%%%%%%%%%
\bibliography{Diff_ref}
\end{document}